\documentstyle[twocolumn,prl,aps,floats]{revtex}
\begin{document}
 
\title
{\Large Anomalous Global Strings and Primordial Magnetic Fields}
       
\author{Robert H. Brandenberger$^1$ and Xinmin Zhang$^2$}

\smallskip
\address{~\\$^1$Department of Physics, Brown University, Providence, RI 02912, USA;
~\\$^2$CCAST (World Laboratory), P.O. Box 8730, Beijing 100080, P.R. China, and Institute of High Energy Physics, Academia Sinica,
Beijing 100039, P.R. China}
 
\maketitle

\begin{abstract}

\noindent We propose a new mechanism for the generation of primordial magnetic fields, making use of the magnetic fields which are induced by anomalous global strings which couple to electromagnetism via Wess-Zumino type interactions. This mechanism can be realized in QCD by utilizing pion strings, global vortices which appear in the linear sigma model which describes physics below the QCD confinement scale. During the chiral symmetry breaking phase transition, pion strings can be produced, thus leading to primordial magnetic fields. We calculate the magnitude and coherence length of these fields. They are seen to depend on the string dynamics. With optimistic assumptions, both the magnitude and coherence scale of the induced magnetic fields can be large enough to explain the seed magnetic fields of greater than $10^{-23}$ Gauss necessary to produce the observed galactic magnetic fields via the galactic dynamo mechanism.
\end{abstract}


\narrowtext

\noindent BROWN-HET-1134, August 1998

\vskip 0.8cm
\section{Introduction}

The origin of galactic magnetic fields is still a mystery. Observations (see e.g. \cite{Beck} for a recent review) indicate that magnetic fields in galaxies have strengths of order $10^{-6}$ Gauss  which are coherent on scales of several kpc. Such fields can be produced by the galactic dynamo mechanism (see e.g. \cite{Subramanian} for a recent study) provided seed magnetic fields of strength of between $10^{-23}$ and $10^{-19}$ Gauss are present initially.

There have been several proposals to generate the required seed magnetic fields (see e.g. \cite{Olinto,Enqvist} for recent reviews). A breaking of electromagnetic gauge invariance in the early Universe$^{\cite{Widrow}}$ or a change in the gravitational coupling$^{\cite{Ratra}}$ are two radical proposals which require changes in basic physical principles. In string cosmology, these effects may naturally arise$^{\cite{Massimo}}$. Phase transitions in the early Universe leads to another class of mechanisms for generating primordial magnetic fields. The random orientation of the fields may produce magnetic fields$^{\cite{Vachaspati}}$ by the Kibble mechanism$^{\cite{Kibble}}$. Although the initial magnitude of the fields can be large on small scales in this scenario, it is hard to generate the required amplitude of the coherent seed fields on large scales$^{\cite{Olinto2}}$. Magnetic field also can be generated in the collision of bubbles formed at a first order phase transition$^{\cite{KV95,AE97}}$, or by dynamical charge separation during the phase transition$^{\cite{Olinto3,Sigl}}$. Once again, it is hard to obtain coherence on large enough scales. Superconducting cosmic strings may lead to magnetic fields via turbulence effects which take place in the sting wakes$^{\cite{Dimopoulos}}$. Astrophysical mechanisms have also been proposed (see e.g. \cite{Rees1,Rees2,Subra2,Subra3}).

In this paper, we propose a new mechanism for the generation of primordial magnetic fields. It is based on the realization$^{\cite{Kaplan}}$ that anomalous global strings couple to electricity and magnetism via an induced $F_{\mu \nu} {\tilde F^{\mu \nu}}$ term in the low energy effective lagrangian and induce magnetic fields (note that such a term also is a crucial feature in magnetic field generation mechanisms$\cite{GFC92,JS97,CF98}$ based on a pseudoscalar coupling to electricity and magnetism). The major advantage of this mechanism is that the coherence scale of the magnetic fields induced by these global vortex lines is set by the length scale $\xi(t)$ of the strings (the typical curvature radius of the strings). In models which admit strings, strings are inevitably produced during the symmetry breaking phase transition in the early Universe$^{\cite{Kibble}}$. Immediately after the phase transition, the string length scale is microscopic. However, the string network rapidly approaches a {\it scaling solution} (see e.g. \cite{VS,HK,RB94} for recent reviews on cosmic strings) during which $\xi(t)$ is proportional to time $t$. In this way, there is a natural way in which large-scale coherent fields can be produced.

A concrete realization of our mechanism exists in QCD. As was shown in \cite{Zhang}, there exists a class of string-like classical solutions of the effective theory (linear sigma model) which describes strong interaction physics below the confinement scale. These solutions are called pion strings. During the chiral symmetry breaking phase transition at a temperature $T_c \sim 100 - 200$MeV, a network of pion strings forms. Pion strings are not topologically stable, and they will hence eventually decay at a temperature $T_d \sim 1$MeV.

We calculate the magnitude and coherence scale of the primordial magnetic fields induced by the anomalous global strings as a function of $T_c$ and $T_d$ and demonstrate that it is possible (but requires streching of some parameters) to use pion strings to generate the required seed magnetic fields of greater than $10^{-23}$ Gauss coherent on comoving scales of a few kpc.

In the following section, we give a brief review of pion strings$^{\cite{Zhang}}$ and discuss the mechanism$^{\cite{Kaplan}}$ by which they generate primordial magnetic fields. In Section 3 we calculate the magnitude and coherence length of the primordial magnetic fields induced by anomalous global vortex lines as a function of $T_c$ and $T_d$. The calculations apply in the general case. We then specialize to the case of pion strings. We conclude the paper with a discussion of further applications of and open issues concerning our general mechanism for magnetic field generation. For most of
the paper, we work in natural units.

\section{Pion Strings and Induced Magnetic Fields} 

In this section we will review the work of \cite{Zhang} in which it was shown that below the chiral symmetry breaking scale, the effective lagrangian of strong interaction physics admits global vortex line solutions, the pion strings. We then use the results of \cite{Kaplan} to demonstrate that these vortex lines generate primordial magnetic fields.

We consider an idealization of QCD with two species of massless
quarks $u$ and $d$. The lagrangian of strong interaction physics is invariant under $SU_L(2) \times SU_R(2)$ chiral transformations
\begin{equation}
\Psi_{L, R} \rightarrow {\rm exp}(-i \vec{\theta}_{L,R} \cdot \vec{\tau} )
         \Psi_{L, R},
\end{equation}
where $\Psi^T_{L,R} = (u, d)_{L,R}$. However this chiral symmetry does
not appear in the low energy particle spectrum since it is spontaneously broken  to the diagonal subgroup $SU_{L+R}(2)$ by the vacuum of QCD via
quark condensate formation.
Consequently, three Goldstone bosons, the pions, appear and the (constituent)
quarks become massive.
At low energy, the spontaneous breaking of chiral symmetry can be described by an effective theory, the linear sigma model, which involves the massless pions ${\vec \pi}$ and a massive $\sigma$ particle.

As usual, we introduce the field
\begin{equation}
\Phi = \sigma \frac{\sigma^0}{2} + i {\vec \pi} \frac{\vec \tau}{2},
\end{equation}
where $\tau^0$ is unity matrix and $\vec\tau$ are the Pauli matrices with 
the normalization condition $Tr(\tau^a \tau^b) =2 \delta^{ab}$. Under
$SU_L(2) \times SU_R(2)$ chiral transformations, $\Phi$ transforms as
\begin{equation}
\Phi \rightarrow L^+ \Phi R.
\end{equation}
The renormalizable effective lagrangian of the linear sigma model is given by$^{\cite{Itzykson}}$
\begin{equation}
{\cal L} = {\cal L}_{\Phi} + {\cal L}_{q},
\end{equation}
where
\begin{equation} \label{eq5}
{\cal L}_{\Phi} = Tr[ {( \partial_\mu \Phi)}^+ \partial^\mu \Phi] - \lambda
{[ Tr( \Phi^+ \Phi) - \frac{f_\pi^2}{2} ]}^2,
\end{equation}
and
\begin{equation} \label{eq6}
{\cal L}_q = {\bar \Psi}_L i \gamma^\mu \partial_\mu \Psi_L + {\bar \Psi}_R
       i \gamma^\mu \partial_\mu \Psi_R -2 g {\bar \Psi}_L \Phi \Psi_R
+ {\it h.c.}.
\end{equation}
During chiral symmetry breaking, the field $\sigma$ takes on a nonvanishing 
vacuum expectation value, which breaks $SU_L(2) \times SU_R(2)$ 
down to $SU_{L+R}(2)$. This results in a massive sigma $\sigma$ $^{\cite{Foot}}$ and three massless Goldstone bosons ${\vec \pi}$, as well as giving a mass $m_q = g f_\pi$ to the constituent quarks. Numerically, $f_\pi \sim 94$ MeV
and $m_{q} \sim 300$ MeV.

In an earlier paper, we studied the classical solutions of this model
and  discovered a class of vortex-like configurations which we refer to as pion string$^{\cite{Zhang}}$. Like
the Z string$^{\cite{Vachaspati2}}$ of the standard electroweak model, the pion string is not topologically stable. Nevertheless, as demonstrated recently in numerical simulations in the case of semilocal strings$^{\cite{Borrill}}$ (which are also not topologically stable) pion strings are expected to be produced during the QCD phase transition in the early Universe (and also in heavy-ion collisions). The strings will subsequently decay.
  
The pion string is a static configuration of the lagrangian ${\cal L}_\Phi$ of Eq. (\ref{eq5}).
To construct these solutions,  we define new fields (following the notation of Ref. \cite{Zhang}) 
\begin{eqnarray}
\phi&=& \frac{ \sigma + i \pi^0}{ \sqrt 2 },\\
\pi^\pm &=& \frac{ \pi^1 \pm i \pi^2 }{ \sqrt 2}.
\end{eqnarray}
The lagrangian ${\cal L}_\Phi$ now can be rewritten as
\begin{equation}
{\cal L} =
{(\partial_\mu \phi )}^* \partial^\mu \phi
+ \partial_\mu \pi^+ \partial^\mu \pi^- - \lambda {( \pi^+ \pi^-
+ \phi^* \phi - \frac{f_\pi^2}{2} )}^2.
\end{equation}
For static configurations, the energy functional corresponding to the above lagrangian is given by
\begin{equation} \label{energy}
E = \int d^3 x 
\left [ {\vec{\bigtriangledown} \phi }^* \vec{\bigtriangledown}\phi
+ \vec{\bigtriangledown}\pi^+ \vec{\bigtriangledown}\pi^- + \lambda
{( \pi^+ \pi^- + \phi^* \phi - \frac{ f_\pi^2}{2} )}^2 \right ].
\end{equation}
The time independent equations of motion are:
\begin{eqnarray}
 {\bigtriangledown}^2 \phi &=&  2 \lambda ( \pi^+ \pi^-
+ \phi^* \phi - \frac{f_\pi^2}{2} ) \phi,\\
 \bigtriangledown^2 \pi^+&=& 2 \lambda ( \pi^+ \pi^- + \phi^* \phi
   - \frac{f_\pi^2}{2} ) \pi^+ .
\end{eqnarray}
The pion string solution extremising the energy functional of Eq. (\ref{energy}) is given by
\begin{eqnarray}
\phi &=& \frac{ f_\pi}{\sqrt 2} \rho (r) e^{i  \theta}, \\
\pi^\pm &=& 0,
\end{eqnarray}
where the coordinates $r$ and
$\theta$ are polar coordinates in the $x-y$ plane (the string is assumed to lie along the z axis), and
$\rho(r)$ satisfies the following boundary conditions
\begin{eqnarray}
r \rightarrow 0, & & \rho(r) \rightarrow 0 ; \\
r \rightarrow \infty, & & \rho(r) \rightarrow 1.
\end{eqnarray}

To study primordial magnetic field generation, we turn on the $U_{em}(1)$
electromagnetic interaction by replacing the derivatives in eqs. (5) and (6)
by covariant derivatives. Since a pion string is made of $\sigma$ and
$\pi^0$ fields, it is neutral under the $U_{em}(1)$ symmetry. However,
the $\pi^0$ will couple to photons via the Adler-Bell-Jackiw anomaly.
In the linear sigma model, the
effective coupling of $\pi^0$ to photons  is obtained from the contribution
of the quark triangle diagram$^{\cite{Pisarski2}}$.
At low energy, only massless particles$^{\cite{Foot2}}$, the pion and photon, are important.
Integrating out the heavy particles, the sigma and the constituent quarks, the
effective lagrangian to leading order is
\begin{equation} \label{low}
{\cal L }_{low}=\frac{f_\pi^2}{4}Tr(\partial_\mu \Sigma^+ \partial^\mu \Sigma)
 - \frac{1}{4}F_{\mu\nu}F^{\mu\nu}
   -\frac{N_c \alpha}{24 \pi} \frac{\pi^0}{f_\pi}
   \epsilon^{\mu\nu\alpha\beta}F_{\mu\nu}F_{\alpha\beta},
\end{equation}
where $N_c = 3$, $\Sigma = exp(i {\vec\tau}\cdot{\vec\pi}/ f_\pi )$, and $\alpha$ is the electromagnetic fine structure constant.

The classical equation of motion for the electromagnetic
field obtained from eq. (\ref{low}) is
\begin{equation} \label{anom}
\partial_\mu F^{\mu\nu} = - \frac{\alpha}{\pi} \partial_\mu (\frac{\pi^0}
{f_\pi})\tilde{F^{\mu\nu}}.
\end{equation}
Similar equations with the anomalous Axion string have been solved
by Kaplan and Manohar in Ref. \cite{Kaplan}. We closely follow their procedure
and hence here only report the results. 

The crucial point is that via the anomaly term in (\ref{anom}) charged zero modes on the string will induce a magnetic field circling the string which falls off less rapidly as a function of the distance from the string than expected classically. Zero mode currents, in turn, are automatically set up by the analog of the Kibble mechanism$^{\cite{Kibble}}$ at the time of the phase transition.
Thus, magnetic fields coherent with the strings are automatically generated during the phase transition.
  
In the background of an infinite string in the $z$ direction,
there are two static solutions which are z-independent
\begin{eqnarray}
E_r = C_{+} r^{-1- {\alpha}/{\pi} } + C_{-}r^{-1 + {\alpha}/{\pi}} , \\
B_{\theta}=C_{+} r^{-1 - {\alpha}/{\pi} } - C_{-} r^{-1 + {\alpha}/{\pi} },
\end{eqnarray}
with the other fields vanishing. 
The coefficients $C_{\pm}$ can be determined by the zero mode current
along the pion string$^{\cite{Kaplan}}$. Note that the electromagnetic fields fall off less fast than expected classically, and that the decay rate depends on the strength of the anomalous coupling.

For the up quark coupled to the pion string,
the Dirac equation can be obtained from ${\cal L}_q$ in (6).
Following from an index theorem$^{\cite{Weinberg}}$ we deduce the existence of a single zero mode solution which is independent of z and t and normalizable in the x-y plane.  In the notation of Ref. \cite{Kaplan}, this solution is
\begin{eqnarray}
u_L = \eta(t, z) exp[ - gf_\pi \int^r_0 \rho(r) dr], \\
u_R = -i \gamma^1 u_L .
\end{eqnarray}
This zero mode corresponds to a fermion travelling in the $- {\hat z}$
direction at the speed of light and generates a zero mode current
\begin{equation}
j_{zero- mode}^\mu = \frac{Q_u N_c n}{2 \pi} f(r)(1, 0, 0, 1),
\end{equation}
where $n$ is the number of quarks per unit length along the pion string,
$Q_u = 2 e /3$ and
\begin{equation}
f(r) = \frac{ exp[ -2g f_\pi \int_0^r \rho(r') dr'] }
{\int^\infty_0 exp[-2 g f_\pi \int_0^{r'} \rho(r") dr"] r' dr' },
\end{equation}
which measures the (normalized) transverse spread of the zero mode current and
falls off$^{\cite{Kaplan}}$ exponentially as $e^{-2 m_{q} r}$.
Down quarks couple to $\phi^* \sim \sigma - i \pi^0$ instead of to
$\phi$. Hence, for down quarks the pion string looks like an anti-vortex$^{\cite{Witten}}$. Consequently, the zero mode 
fermion travels in the opposite direction. Since $Q_u = \frac{2}{3} e$ and
$Q_d = - \frac{1}{3} e$, the net zero mode currect flows in the 
$- {\hat z}$ direction,
\begin{equation}
j^\mu = N_c \frac{e n}{2 \pi} f(r)(1, 0, 0, 1).
\end{equation}
This allows the determination$^{\cite{Kaplan}}$ of the coefficients $C_+$ and $C_-$, with the result $C_{+}=0$ and $C_{-}= N_c \frac{e n}{2 \pi} r_0^{- \alpha / \pi}$, where
that $r_0 \geq {( 2 m_{q} )}^{-1}$. Thus we have
\begin{equation}
E_r = - B_{\theta} = N_c \frac{e n} {2 \pi} {( 2 m_{q} )}^{\alpha / \pi}
   r^{-1 + \alpha / \pi} .
\end{equation}

\section{Seed Magnetic Field from Global Strings}

In this section we estimate the magnitude of the magnetic field generated by global anomalous strings which are coherent on galactic scales. The two important scales are temperature $T_c$ when the symmetry breaking phase transition which gives rise to the strings occurs, and the temperature $T_d$ when the string network decays (when the strings become unstable). Plasma effects will, for the moment, be neglected. We will briefly discuss these effects in the discussion section.

Generalizing the discussion of the previous section of magnetic fields from pion strings to the case of general anomalous global strings, we conclude that the coherent magnetic fields induced by the strings at a distance $r$ from the string have the magnitude
\begin{equation}
B \, = \, N_c {{e n} \over {2 \pi}} ({r \over {r_0}})^{\alpha / \pi} {1 \over r} \, ,
\end{equation}
where $n$ is the number density of charge carriers on the string, $r_0$ is the width of the string and $\alpha \ll 1$. By dimensional analysis, at the time $t_c$ when the strings form, their width will be $r_0 \sim T_c^{-1}$, and the number density of charge carriers is $n \sim T_c$. For our general study, we
take $\alpha$ to be an effective anomalous coupling. In our specific model
based on the chiral phase transition of QCD, it is the electromagnetic coupling
constant. However, it could be larger than that for a model with a big
coeefficient of the Wess-Zumino term.

The evolution of the string network in the time interval during which the Universe cools from $T_c$ to $T_d$ is complicated. Initially, the strings have a typical curvature radius and separation of $\xi(T_c)$ (``correlation length") which then increases rapidly and eventually approaches a scaling solution in which $\xi(t) \sim t$. During this evolution, the charge density is diluted as the strings stretch (by total charge conservation), but the mergers of small string segments into larger ones leads to a buildup of charge which can be modelled as a random walk superposition of the charges of the individual segments. Hence, the charge carrier density $n(t_d)$ of a string at time $t_d$ when the correlation length is $\xi(t_d)$ is
\begin{equation} \label{dilution}
n(t_d) \, \sim \, {{\xi(t_c)} \over {\xi(t_d)}} \bigl({{\xi(t_d)} \over {\xi(t_c)}}\bigr)^{1/2} n(t_c) \, ,
\end{equation}
where the first factor on the r.h.s. comes from the stretching, and the second from the random walk superposition.

If the initial separation of strings is microscopic, then their evolution at early times will be friction-dominated and
\begin{equation} \label{scale}
\xi(t) \, \sim \, t^p
\end{equation}
where$^{\cite{friction}}$ $p = 5/4$ or$^{\cite{BR98}}$ $p = 3/2$ until $\xi(t)$ becomes comparable to the Hubble radius $t$. At late times, the correlation length scales as $t$ (i.e. $p = 1$ in (\ref{scale})). We assume that the strings decay during the radiation-dominated epoch. In this case it follows from (\ref{dilution}) that
\begin{equation}
n(t_d) \, \sim \, \bigl({{T_d} \over {T_c}}\bigr)^p n(t_c) \, .
\end{equation}

Combining the above results, it follows that
\begin{eqnarray}
B(T_d) \, & \sim & \, N_c {e \over {2 \pi}} {{T_c} \over r} \bigl({{T_d} \over {T_c}}\bigr)^p (r T_c)^{\alpha / \pi} \\ 
& = & \, N_c {e \over {2 \pi}} {{T_c} \over {1 {\rm GeV}}} r_m^{-1} \bigl({{T_d} \over {T_c}}\bigr)^p  (r T_c)^{\alpha / \pi} {{\rm{GeV}} \over {\rm{m}}} \, ,
\end{eqnarray}
where $r_m$ is the distance in meters. Converting from natural to MKSA units (and remembering to insert the factors of $c$ and $\mu_0$) this gives
\begin{equation}
B(T_d) \, \sim \, 10^5 {{T_c} \over {1 {\rm GeV}}} r_m^{-1} \bigl({{T_d} \over {T_c}}\bigr)^p  (r T_c)^{\alpha / \pi} {\rm Gauss} \, .
\end{equation}

Between string decay and the present time, we assume that the magnetic field lines are fixed in comoving coordinates. By flux conservation, it hence follows that the magnitude of $B$ at some fixed comoving point scales as $z(t)^2$, where $z(t)$ is the cosmological redshift. We are interested in comoving distances from the string which are of galactic scales, i.e. of the order kpc ($10^{19}$m). At the time $t_d$, the corresponding distance $r$ was smaller by a factor of $z(t_d)^{-1}$. Hence
\begin{equation} \label{final}
B(t_0) \, \sim \, 10^{-14} {{T_c} \over {1 {\rm GeV}}} r_{\rm kpc}^{-1} \bigl({{T_d} \over {T_c}}\bigr)^p  (r T_c)^{\alpha / \pi} \bigl({{T_0} \over {T_d}}\bigr) {\rm Gauss} \, ,
\end{equation}
where $t_0 (T_0)$ denote the present time (temperature), and $r_{\rm kpc}$ is the present distance from the original comoving location of the string expressed in kpc. Note that $r$ is the physical distance at $T_d$.

In the case of the pion string, we make the assumption that $\xi(t) \sim t$ for all times. In contrast, topologically stable strings formed at the QCD scale would be formed with a microscopic $\xi(t_c)$ and would remain in the friction-dominated period with $\xi(t) \ll t$ until the present time. However, since pion strings are not topologically stable, their formation probability will be smaller and it is not unreasonable to take the causality bound $t$ as an estimate for the string separation. Given this assumption, we can estimate the resulting magnetic field strength by setting $T_c \sim 1{\rm GeV}$, $T_d \sim 1{\rm MeV}$ and $p = 1$ in (\ref{final}), thus obtaining
\begin{equation} \label{pionresult}
B(t_0) \, \sim \, 10^{-26} (r T_c)^{\alpha / \pi} {\rm Gauss} \, .
\end{equation}
Since $r$ is a cosmological distance and $T_c^{-1}$ is microscopic, the enhancement factor (the second term on the r.h.s. of (\ref{pionresult})) can be large (if the effective coupling $\alpha$ is large) and may boost the magnitude of $B(t_0)$ to a value large enough to seed the galactic dynamo effect.

We must also worry about the coherence scale of the induced magnetic field at the present time. Once again, making use of the reasonable assumption that after $T_d$ the field lines are frozen in comoving coordinates, the coherence length is $\xi(t_d)_c$, where the subscript indicates comoving coordinates, with
\begin{equation}
\xi(t_d)_c \, = \, \beta t_d z(t_d) \, = \, \beta 10^{-2} {\rm kpc} T_d[{\rm MeV}]^{-1} \, ,
\end{equation}
where $\beta$ is a constant which for scaling strings is of the order 1.
On scales larger than $\xi(t_d)_c$, the fields add up incoherently as a random walk, yielding for the average field
\begin{equation}
{\bar B} \, = \, \bigl({1 \over N}\bigr)^{1/2} B(t_0) \, ,
\end{equation}
where
\begin{equation} \label{number}
N \, = \, \bigl( {d \over {\xi(t_d)_c}}\bigr)^2 \, = \, d[{\rm kpc}]^2 \beta^{-2} 10^4 T_d[{\rm MeV}]^2 \, ,
\end{equation}
where $d$ is the scale on which we want to calculate the coherent magnetic field. For the values of the pion string chosen above, ${\bar B}$ is reduced compared to (\ref{pionresult}) by about two orders of magnitude. Note, however, that the magnetic field lines may actually not be frozen in comoving coordinates$^{\cite{DD}}$, in which case the suppression factor associated with the $N$ of (\ref{number}) would be reduced.

Thus, we conclude that although the global string mechanism discussed in this paper can easily explain how QCD-scale strings can generate primordial magnetic fields which are coherent on the required scales, one requires either a large value of the anomalous coupling $\alpha$ in order that the amplitude is sufficient for seeding a galactic dynamo, or else one needs to invoke nonlinear hydrodynamical effects$^{\cite{BE1,BE2}}$ which may increase the magnitude of the magnetic fields by several orders of magnitude.

\section{Discussion}  

In this paper we have discussed the cosmological magnetic fields induced by global anomalous strings. Due to the existence of charged zero modes on the string, coherent magnetic fields are generated when the strings form at a temperature $T_c$ and evolve thereafter. We have pointed out that the stretching of the string network provides a natural mechanism of increasing the coherence length of the magnetic fields to scales which could be important for cosmology. 
 
We have focused our attention on a particular example, pion strings which form at the QCD phase transition. Pion strings are unstable and hence the network of strings will decay at a temperature $T_d \ll T_c$. We have calculated the amplitude and coherence length of the magnetic fields as a function of $T_c$, $T_d$ and of the effective coupling $\alpha$ to the anomaly. This calculation is applicable to general theories with global anomalous strings, in particular to axion strings.

The typical coherence scale of the induced magnetic fields is determined by the decay temperature $T_d$. Provided that $T_d < 10^{-2} {\rm MeV}$ then the coherence scale of the fields induced by a single string at $T_d$ is sufficiently large to explain the length scale of coherent galactic magnetic fields (assuming that the string network is scaling at $T_d$).

The amplitude of the induced magnetic fields depends sensitively on the history of the string network. We obtain the largest amplitude if we assume that the strings were scaling throughout their history (i.e. $p = 1$ in (\ref{scale})).
In this case, the amplitude is independent of $T_d$ and depends on $T_c$ only via factor $(r T_c)^{\alpha / \pi}$ which comes from the crucial fact that the fields fall off less fast as a function of the distance from the string than would be expected from classical considerations. Neglecting this factor, we find that the amplitude is too small by a few orders of magnitude to explain the required seed fields for the galactic dynamo. However, by taking this factor into account, it may be possible to generate sufficiently large fields. If
$T_c$ is the QCD scale and $\alpha$ the electromagnetic fine structure constant, the enhancement factor is too small (only order unity). However, if the effective coupling to the anomaly is large or if $T_c$ is very large, such as for axion strings, then
the fields generated by the mechanism discussed in this paper will be sufficiently large to explain the seed fields for the galactic dynamo. Even
without the enhancement factor coming from a large value of $\alpha$, the magnetic fields may be sufficiently large if nonlinear hydrodynamical effects$^{\cite{BE1,BE2}}$ are taken into account.

Plasma effects have been neglected in this paper. However, it is known that
a magnetized plasma will resist the expansion of magnetic fields in comoving coordinates (see e.g. \cite{Field} and references therein). The time scale $\tau(l)$ required for magnetic fields to diffuse a physical distance $l$ in a plasma of temperature $T$ is 
\begin{equation}
\tau = \sigma l^2 \, ,
\end{equation}
where $\sigma$ is the diffusion constant. By dimensional analysis (see also \cite{Massimo}), we expect $\sigma \sim T$. Hence, it follows that the time scale for a magnetic field to diffuse a distance $\xi(T_d) \sim t_d$ at a plasma temperature $T_d$ is much larger than the Hubble time at the corresponding time $t_d$, which implies that before recombination, the magnetic fields produced by the strings will not look like the vacuum configurations on a scale of $t_d$  However, after recombination the Universe will be electrically neutral, and the resistance to the expansion of the magnetic fields will disappear. Provided that the plasma effects do not destroy the magnetic fields set up at the time $t_d$ by the strings, the fields will rapidly approach their vacuum distribution calculated in the previous section.   

Since there are generically charged zero modes for superconducting cosmic strings, the mechanism for magnetic field generation discussed here may also apply to such strings (a very similar mechanism has been proposed in \cite{Martins}). However, one must be careful to consider the cosmological constraints on the amplitude of zero modes on superconducting cosmic strings$^{\cite{BDMT}}$.

\begin{center}{\Large  Acknowledgements}\end{center}

We thank Tao Huang, A.-C. Davis and G. Field for discussions. R.B. wishes to thank the Institute of High Energy Physics of the Academia Sinica for its generous hospitality during a visit when this project was started. This work 
was supported in part by National Natural Science Foundation 
of China and by the U.S. Department of Energy under Contract DE-FG02-91ER40688, TASK A.


\begin{thebibliography}{10}
\bibitem{Foot}
 The identification of the $\sigma$
particle is highly controversial. In the Particle Data Tables$^{\cite{PDB}}$ it is listed as a very broad meson with $m_{\sigma} \sim 400 - 1200$ MeV and
$\Gamma_{\sigma} \sim 600 - 1200$ MeV. In nuclear physics, the $\sigma$ particle
plays an important role and there are many recent works devoted to searching
for it in $\pi \pi$ scattering$^{\cite{Kozlov}}$. In recent years, 
the $SU_L(2) \times SU_R(2) \sim O(4)$ linear sigma-model has been
often used as a model in studies of the physics associated with the 
chiral phase transition$^{\cite{Pisarski}}$ and of the disoriented chiral condensates$^{\cite{Rajagopal}}$ in heavy ion collisions.
\end{thebibliography}
\end{document}